\documentclass[twocolumn,showpacs,aps,prl, superscriptaddress]{revtex4-1}
\usepackage{graphicx}
\usepackage{color}\usepackage{enumerate}
\usepackage{amsmath}\usepackage{amssymb}\usepackage{stmaryrd} \usepackage{mathrsfs}
\usepackage{multirow}
\usepackage{float}
\usepackage{longtable,booktabs}
\usepackage{subfigure}
\usepackage{dsfont}
\usepackage{amssymb}
\usepackage[pdftex,colorlinks=true,citecolor=blue,linkcolor=blue,urlcolor=blue]{hyperref}

\begin{document}
\title{Coexistence of Localized and Extended States in Disordered Systems}
\author{Yi-Xin Xiao$^1$, Zhao-Qing Zhang$^1$ and C. T. Chan*}
\affiliation{Department of Physics, Hong Kong University of Science and Technology, Clear Water Bay, Kowloon, Hong Kong}
\begin{abstract}
It is commonly believed that Anderson localized states and extended states do not coexist at the same energy. Here we propose a simple mechanism to achieve the coexistence of localized and extended states in a band in a class of disordered quasi-1D and quasi-2D systems. The systems are partially disordered in a way that a band of extended states always exists, not affected by the randomness, whereas the states in all other bands become localized. The extended states can overlap with the localized states both in energy and in space, achieving the aforementioned coexistence. We demonstrate such coexistence in disordered multi-chain and multi-layer systems. 
\end{abstract}

\maketitle
Disorder plays an important role in many fields of physics. It is well known that interference between multiply scattered waves can disrupt the transport of electrons and classical waves in disordered media and make the transport diffusive \cite{sheng_introduction_2006, lagendijk_fifty_2009, abrahams_50_2010}. In 1958, Anderson predicted that disorder can even completely stop diffusion \cite{anderson_absence_1958}. The phenomenon, now called Anderson localization \cite{anderson_absence_1958, sheng_introduction_2006, lagendijk_fifty_2009, abrahams_50_2010}, has profound influence on our understanding of wave transport and has been extensively studied in the past few decades \cite{de_raedt_transverse_1989, dunlap_absence_1990, ye_observation_1992, wiersma_localization_1997, chabanov_statistical_2000, storzer_observation_2006, topolancik_experimental_2007, schwartz_transport_2007, roati_anderson_2008,  evers_anderson_2008, lahini_anderson_2008, hu_localization_2008, chabe_experimental_2008, billy_direct_2008, lahini_observation_2009, kondov_three-dimensional_2011, jendrzejewski_three-dimensional_2012, segev_anderson_2013, hsieh_photon_2015}. A widely-accepted feature of Anderson localization is that all states are localized in one-dimensional (1D) and two-dimensional (2D) disordered systems due to coherent backscattering effects \cite{abrahams_scaling_1979, sheng_introduction_2006}, while three-dimensional (3D) random media have a mobility edge \cite{mott_metal-insulator_1990, economou_localization_1970, licciardello_conductivity_1975, john_electromagnetic_1984, kuhl_experimental_2000, delande_mobility_2014, semeghini_measurement_2015} that separates localized states from extended states  \cite{sheng_scalar-wave_1986, li_anisotropic_1989}. These features were predicted by both the scaling theory \cite{abrahams_scaling_1979} and the self-consistent theory \cite{vollhardt_scaling_1982} of Anderson localization. The concept of mobility edge was first introduced by Mott who argued that if a localized state and an extended state are infinitely close in energy, they will get hybridized by an infinitesimal change of the system and turn into two extended states \cite{mott_metal-insulator_1990, economou_localization_1970}. Mobility edge has also been found in one-dimensional incommensurate quasi-periodic models \cite{biddle_predicted_2010, ganeshan_nearest_2015}. Recently, Anderson localization has also been observed in 3D non-interacting cold atoms under the influence of laser speckle potentials \cite{semeghini_measurement_2015, pasek_anderson_2017}. The mobility edge was found to obey a scaling behavior, independent of the speckle geometry \cite{semeghini_measurement_2015, pasek_anderson_2017}. \\
\indent The argument for the spectral separation of localized and extended states also applies to a similar question raised by Von Neumann and Wigner in 1929 \cite{von_neuman_j._1929}, namely whether a bound state can exist in the continuum of extended states (BIC). In the past few years, BICs have been achieved by several different mechanisms \cite{friedrich_interfering_1985, hsu_observation_2013, plotnik_experimental_2011, mur-petit_chiral_2014, zhen_topological_2014, xiao_topological_2017, bulgakov_topological_2017, hsu_bound_2016} in various ordered systems. The challenging question is to find systems which support spectral coexistence of Anderson localized states and extended states. The task is very different from searching BICs as the media are disordered and Anderson localized states can also form a continuous spectrum \cite{economou_greens_2006}. Until now, there is no guiding principle for the creation of the coexistence of Anderson localized states and extended states in a band. Some attempts were made in the past to achieve this goal. It was shown that an inhomogeneous trap combined with a disordered potential can lead to the coexistence of localized and extended states in energy, but spatially they do not overlap as the inhomogeneous trap segregates two types of states and suppresses hybridization \cite{pezze_localized_2011}. The coexistence was also found in Ref.  \cite{rodriguez_controlled_2012}, however, complicated engineering of the hopping parameters and on-site energies is needed. Thus, the challenging problem remains to be resolved whether it is possible to have simple disordered systems that support the coexistence of Anderson localized states and extended states in a band. \\
\indent In this work, we find such coexistence in a class of quasi-1D and -2D systems. Our mechanism of creating the coexistence is to partially randomize the systems in a way that the Hilbert space can be partitioned into two subspaces so that the states in one subspace are unaffected by the presence of randomness and hence they can be extended, whereas wavefunctions in all other bands become localized by randomness. The coexistence of two types of states in a band is naturally achieved. We explicitly demonstrate the coexistence in multi-chain and multi-layer systems. \\
\indent We first numerically demonstrate such coexistence in a multi-chain system described by a nearest-neighbor tight-binding Hamiltonian. The system consists of $2N+1$ ($=51$) coupled chains placed in the $x$ direction, as depicted in Fig. 1 (a). In the absence of disorder, the system is periodic in the $x$ direction and all the chains are identical. There are $2N+1$ sites per unit cell. The lattice constant is $a=1$, and all on-site energies and hopping parameters are taken to be zero and a constant $t$, respectively. The band structure of the multi-chain system, as shown in Fig. 1(b), comprises $2N+1$ bands. The central band marked by the blue curve has exactly the same dispersion relation $E/t=2 \cos⁡(ka)$ as that of an isolated chain. Now we partially randomize the system by adding random on-site energies $\varepsilon_i$ to every second chain, namely the red sites in Fig. 1(a). Now $N$ out of $2N+1$ chains are disordered. To study the Anderson localization properties of such a system, we truncate each chain in the $x$ direction to $M$ ($=60$) sites and numerically calculate the eigen-functions and the participation ratio ($PR$) of each eigenstate, which is defined as $PR=(\sum_n |c_n |^2 )^2/(\sum_n |c_n |^4 )$, where $c_n$ are the components of an eigenstate $| \varphi \rangle=\sum_n c_n | n\rangle $ and $| n \rangle$⟩ denotes the atomic orbitals. For the case of uniform randomness $\varepsilon_i/t  \in [-5,5]$, the calculated energy spectrum and the participation ratios of all eigenstates are plotted in increasing energy in Fig. 1 (c) and (d), respectively. Fig. 1(d) shows clearly a stark contrast of two types of states: Type 1 contains $M$ ($=60$) states with large values of $PR$ (marked by blue dots) and Type 2 contains $2N \times M$ ($=3000$) states with much smaller $PR$ (marked by red dots). These two types of states overlap spectrally in the band of extended states. We will show analytically later that Type 1 states are extended states having exactly the same spectrum as that of a single isolated chain, i.e., the blue curve in Fig. 1(b), unaffected by the randomness, and Type 2 states are Anderson localized states. We show the wavefunctions of two types of states at the same chosen energy in Fig. 1(e) and (f), which clearly demonstrate their spatial overlap. Thus, we have demonstrated the coexistence of Anderson localized and extended states in a band in such systems. It should be stressed that although the extended states constitute an insignificant portion, i.e. $1/(2N+1)$  , of the whole space of eigenstates in the thermodynamic limit $N \to \infty$, nevertheless they occupy a significant portion of the entire spectrum as shown in Fig. 1(b) and (c). 
\begin{figure}
\centering
\includegraphics[width=0.95\columnwidth]{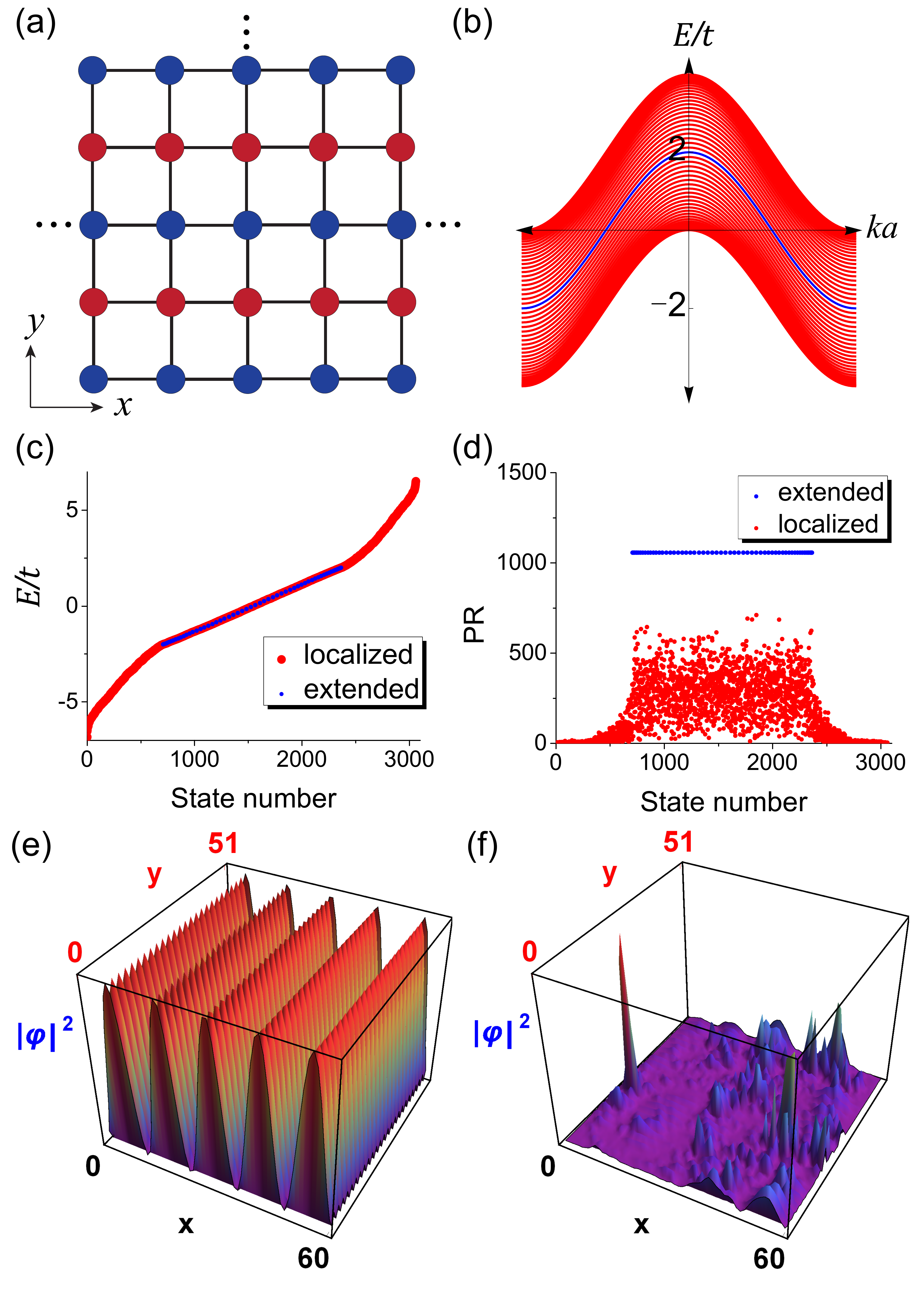}
\caption{(a) shows a square lattice comprising multiple chains extending in the $x$ direction. (b) The band structure with one band in blue color characterized by the same dispersion relation as that of an isolated chain. (c) The energy spectrum showing the spectral coexistence of extended states and localized states in the presence of onsite disorder on alternating (red) chains. (d) The participation ratios of all the eigenstates. (e) and (f) show two states with the same eigen-energy $E/t=1.93$, one is extended and the other is localized.}
\label{fig:1}
\end{figure}

The surprising coexistence phenomenon in such a simple disordered system is in fact the consequence of  
a solid robust mechanism that can be used to realize the coexistence of localized and extended states in a class of systems in all dimensions. Basically, such systems consist of multiple identical copies of $A$ subsystem with  adjacent copies separated and coupled by an intermediate $B$ subsystem. The $A$ and $B$ subsystems can be chains or layers. In the absence of randomness, $B$ subsystems are also identical. A simple example has been shown in Fig. 1(a), in which $A$ and $B$ subsystems are identical chains but denoted by blue and red, respectively. The Hamiltonian of a system containing $N+1$ copies of $A$ and $N$ copies of $B$ can be expressed as
\begin{eqnarray}
\label{eq:Hsys}
H_{sys}=
\left( \begin{array}{ccccc}
H_A & 0 & \cdots & 0 & T_1^\dagger  \\
0 & H_A & \cdots & 0 & T_2^\dagger \\
\vdots & \vdots & \ddots & \vdots & \vdots  \\
0 & 0 & \cdots & H_A & T_{N+1}^\dagger \\
T_1 & T_2 & \cdots & T_{N+1} & \mathcal{H}_B \\
\end{array} \right), \\
\mathcal{H}_B=diag(H_B,H_B,\cdots,H_B ), 
\end{eqnarray}
where $H_A$ and $H_B$ are, respectively, the Hamiltonians of \textit{A} and \textit{B} subsystems. $\mathcal{H}_B$ contains $N$ copies of $H_B$. If we assume that all subsystems are finite, each with $M$ sites, the dimensions of $H_A$ and $H_B$ are $M\times M$. The dimensions of $H_{sys}$  and $\mathcal{H}_B$ in Eq. \eqref{eq:Hsys} become $(2N+1)M \times (2N+1)M$ and $NM \times NM$, respectively. The block matrices $T_i$ denote the couplings between \textit{A} and \textit{B} and have dimensions $M\times NM$. The explicit form of $H_{sys}$ in Eq. \eqref{eq:Hsys} for small values of $N$ and $M$ can be found in Ref.  \cite{SM}. It will be shown analytically below that the coupled system $H_{sys}$ always contains a set of eigenvalues which are the same as those of an isolated  \textit{A} subsystem, similar to the blue band in Fig. 1(b). This implies that the Hamiltonian can be block diagonalized to contain one isolated block $H_A$, namely 
\begin{eqnarray}
\label{eq:unitary}
Q^{-1}H_{sys}Q=
\left( \begin{array}{cc}
H_A & 0  \\
0 & \mathcal{H}'  \\
\end{array} \right), 
\end{eqnarray}
where $Q$ is a unitary matrix and $\mathcal{H}'$ is the other block with dimensions $2NM \times 2NM$. From Eq. \eqref{eq:unitary}, it is easy to see that the Hilbert space of the system is partitioned into two subspaces. Now we introduce randomness into every \textit{B} subsystem, which randomizes the block $\mathcal{H}'$. Since the subspace associated with $H_A$ is not affected by the disorder, all eigenstates in this subspace are extended. The randomness in the other subspace represented by $\mathcal{H}'$ gives rise to localized states. It is shown in ref. \cite{SM} that each extended eigenstate in $H_A$ subspace is a direct sum of all eigenvectors of the $N+1$ \textit{A} subsystems. Since the localized states in $\mathcal{H}'$ subspace also involve atomic orbitals of the $A$ subsystem, the spatial coexistence of two types of states naturally occur. The spectral overlap can always be achieved by adjusting the randomness. We thus have a very general mechanism that assures the coexistence of Anderson localized states and extended states in a class of quasi-1D and -2D systems.
\begin{figure}
\centering
\includegraphics[width=0.75\columnwidth]{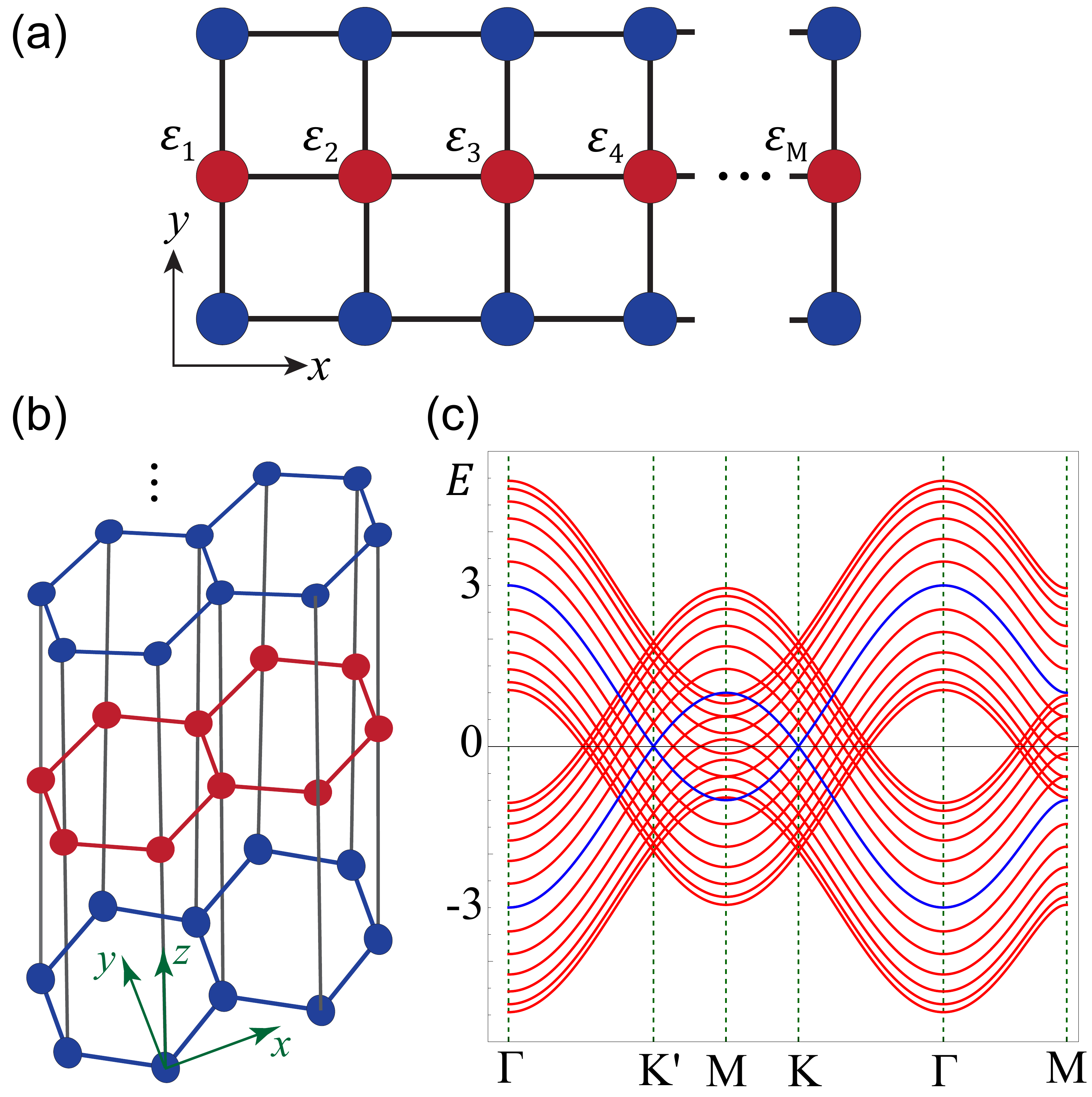}
\caption{(a) shows a three-chain system, each chain contains $M$ sites and on-site disorder is introduced to the middle chain. (b) shows a multilayer system comprising multiple $AA$-stacked honeycomb-lattice layers. (c) The band structure of the multilayer system.}
\label{fig:2}
\end{figure}

To be more explicit, we consider a minimal model of three coupled chains, each truncated to contain $M$ sites, where random on-site energies $\varepsilon_i$ are introduced in the middle chain as shown in Fig. 2(a). The Hamiltonian of the system becomes 
\begin{eqnarray}
\label{eq:Hm}
H_m=
\left( \begin{array}{ccc}
H_A & 0 & T_1^\dagger \\
0 & H_A & T_2^\dagger \\
T_1 & T_2 & H_B
\end{array} \right), 
\end{eqnarray}
where all the matrix elements are block matrices with dimensions $M \times M$. $H_A$ and $H_B$ denote, respectively, the Hamiltonians of the blue and red chains. The inter-chain couplings are represented by $T_1=t I_M$ and $T_2=t I_M$, where $I_M$ denotes an identity matrix. We assume that the single-chain Hamiltonian $H_A$ satisfies the eigen-equation $H_A P=P\Lambda$, where $\Lambda=diag(\lambda_1, \lambda_2,\cdots, \lambda_M )$ is a diagonal matrix and $\lambda_i$'s are the eigenvalues of $H_A$, and $P=(\varphi_1, \varphi_2, \cdots, \varphi_M )$ comprises $M$ columns of eigenvectors. A similarity transformation $H_S=X^{-1} H_m X$ with $X=diag(P, P, I_M )$ can be applied so that
\begin{eqnarray}
\label{eq:HS}
H_S=
\left( \begin{array}{ccc}
\Lambda & 0 & P^{-1} T_1^\dagger \\
0 & \Lambda & P^{-1} T_2^\dagger \\
T_1 P & T_2 P & H_B
\end{array} \right). 
\end{eqnarray}
We note that when inter-chain couplings are absent, namely $T_1=T_2=0$, there are two degenerate sets of eigenvalues $\lambda_i, i=1,\cdots,M$. The presence of inter-chain couplings will split the degeneracy. However, there always exists a set of eigenvalues $\lambda_i$ unaffected by the couplings. This can be seen as follows. For each degenerate pair of eigenvalue $\lambda_i$  of $H_A$, the effects due to $H_B$ can be described by a simplified version of Eq. \eqref{eq:HS}, i.e., 
\begin{eqnarray}
\label{eq:single}
\left( \begin{array}{ccc}
\lambda_i & 0 & t \varphi_i^\dagger \\
0 & \lambda_i & t \varphi_i^\dagger \\
t \varphi_i & t \varphi_i & H_B
\end{array} \right). 
\end{eqnarray}
Since the rank of the coupling block $[t \varphi_i, t \varphi_i]$ is $1$, due to the rank-nullity theorem \cite{meyer_matrix_2000}, $\lambda_i$ remains to be an eigenvalue of $H_S$, thus also of $H_m$, independent of $H_B$. Applying the same argument to all $\lambda_i$, we have shown that a set of eigenvalues of $H_A$ remains and, therefore, $H_m$ can always be block diagonalized into the following form,
\begin{eqnarray}
\label{eq:HBD}
H_{BD}=
\left( \begin{array}{cc}
H_A & 0  \\
0 & \mathcal{H}'  \\
\end{array} \right)
=
\left( \begin{array}{ccc}
H_A & 0 & 0 \\
0 & H_A & R^\dagger \\
0 & R      & H_B
\end{array} \right), 
\end{eqnarray}
by a unitary transformation $H_{BD}=Q^{-1} H_m Q$. Concrete examples with explicit block diagonalization can be found in ref. \cite{SM}. The above block diagonalization means a particular separation of Hilbert space into two subspaces. We call the subspace spanned by the eigenvectors corresponding to the invariant eigenvalues $\lambda_i$ as the invariant subspace. For each eigenvalue $\lambda_i$, the associated eigenvector $\Phi_i$ can be expressed in terms of the direct sum of three parts corresponding to the three chains, namely
\begin{eqnarray}
\label{eq:eigenstate}
\Phi_i=-\frac{1}{\sqrt{2}} \varphi^A_{1,i} \oplus \frac{1}{\sqrt{2}} \varphi^A_{2,i} \oplus \varphi^B,
\end{eqnarray}
where $\varphi^A_{1,i}$ and $\varphi^A_{2,i}$ are the normalized eigenvectors of the two $A$ (blue) chains and both correspond to $\lambda_i$. And $\varphi^B$ is a zero vector with $M$ components. That is to say, the anti-symmetric ``combination'' of the eigenvectors $\varphi^A_{1,i}$ and $\varphi^A_{2,i}$ of the two separate blue chains constitutes an eigenvector $\Phi_i$ for the whole coupled-chain system, which has odd parity, conforming to the mirror symmetry in the \textit{y} direction, and vanishes at the sites in the middle chain. The fact that the degeneracy of $\lambda_i$ (in the absence of inter-chain couplings) outnumbers the coupling channels brought by inter-chain couplings guarantees $\lambda_i$ to be an eigenvalue for the coupled-chain system. Note such a block diagonalization is valid for more general configurations of inter-chain couplings, such as including next-nearest-neighbor hoppings \cite{SM}. Since the invariant subspace does not involve the sites at the middle chain, the disorder (both diagonal and off-diagonal) introduced into the middle chain will not affect the invariant subspace. Consequently, states in the invariant subspace will remain extended, whereas all other eigenstates corresponding to the $\mathcal{H}'$ subspace become localized due to the presence of disorder. The spectral coexistence can always be achieved by adjusting the randomness. Since both subspaces involve the atomic orbitals at the $A$ chains, the extended states and localized states will surely also coexist spatially \cite{SM}.  
% Similar ideas can be used to achieve topological BICs \cite{xiao_topological_2017} and exotic topological phase \cite{bergholtz_topology_2015}.

We can generalize the system from three chains to $2N+1$ chains with $N+1$ identical $A$ chains separated and coupled by another $N$ $B$ chains. Following the similar procedure, it can be shown that the Hilbert space can be split into two subspaces with an invariant subspace spanned by $M$ eigenvectors having the form $\Phi_i=(\bigoplus_{n=1}^{N+1} c_n \varphi^A_{i,n}) \oplus \varphi^B$ ($i=1,\cdots,M$), each corresponds to the eigenvalue $\lambda_i$ of the whole coupled-chain system. Here $\varphi^A_{i,n}$ is the normalized eigenvector for the $n$-th $A$ chain with eigenvalue $\lambda_i$ and $\varphi^B$ is a $NM$-component zero vector. The key is that the identical $A$ chains outnumbers the $B$ chains by $1$ so that eigenvector components on each adjacent pair of $A$ chains add destructively and vanish at the intermediate $B$ chain \cite{SM}. There is always an invariant subspace and all the eigenstates therein are extended, whatever disorder is added on the $B$ chains. The coexistence of extended states and localized states shown in Fig. 1 validates the above analysis.
\begin{figure}
\centering
\includegraphics[width=\columnwidth]{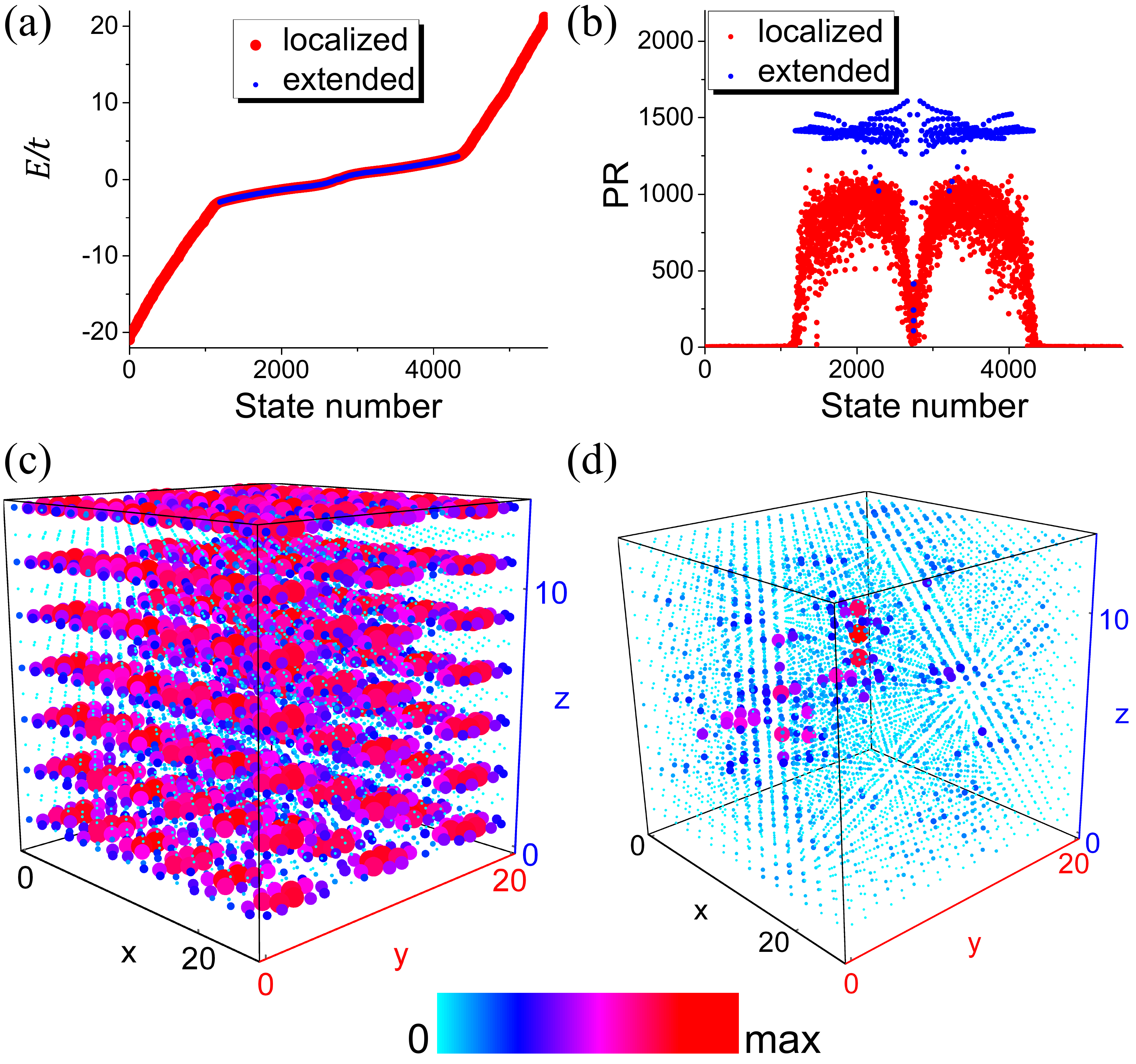}
\caption{(a) shows the coexistence of extended states and localized states in the energy spectrum of a system consisting of multiple honeycomb-lattice layers, where diagonal disorder is added to alternating layers. (b) The participation ratios of all the eigenstates. (c) and (d) show two states with the same eigen-energy $E/t=2.51$, one is extended and the other is localized.}
\label{fig:3}
\end{figure}

Noticing the generality of the above analysis, it is quite obvious that the particular separation of Hilbert space is not limited to quasi-1D systems, but can also be applied to quasi-2D multilayer systems constructed similarly to achieve the spectral and spatial coexistence in coupled-layer systems. As an example, we consider a system comprising $2N+1$ ($N=6$) $AA$-stacked honeycomb-lattice layers as shown in Fig. 2(b). Assuming the system is periodic in the $x$ and $y$ directions, we can compute its band structure. Since the honeycomb lattice is a two-band model, there should be totally $2(2N+1)$ ($=26$) bands as shown in Fig. 2(c), in which two invariant bands are denoted by blue curves. We now introduce uniform diagonal disorder $\varepsilon_i \in [-20, 20]$ to every $B$ (red) layer and the system is truncated to contain $420$ sites per layer. The energy spectrum and participation ratios are shown in Fig. 3 (a) and (b). They clearly show the spectral coexistence of the two different types of states in a band. A few eigenstates near zero energy have low participation ratios but are depicted by blue dots (marked as extended states), because they are edge states localized on the zigzag edges of ordered (\textit{A}) layers and are actually from the invariant subspace. To further demonstrate the spectral coexistence, the absolute values of wavefunctions $|\Psi (x,y,z)|$ of two states at the same energy are denoted by colored dots at $(x,y,z)$, as shown in Fig. 3 (c) and (d). For better visualization, the sizes of the dots are scaled to be proportional to $|\Psi (x,y,z)|$. The two states show markedly different nature: one is extended, and the other is localized. Their spatial coexistence is clearly seen. 

In short, we proposed a method to create the coexistence of localized and extended states in a band in a class of systems, in which the Hilbert space can be partitioned in a way that the disorder affects only one subspace causing localization, while the states in the other subspace remain extended. The coexistence is demonstrated explicitly in both multi-chain and multi-layer systems. We want to emphasize that the method we propose is very general. It is not limited to the multi-chain and multi-layer systems demonstrated here. It applies universally to any similar structures as long as the identical copies of $A$ subsystem outnumber that of $B$ subsystem so that there is a particular subspace originated from $A$ subsystems without involving degrees of freedom of the $B$ subsystems, namely sites on the $B$ subsystems. Such systems can be experimentally realized using coupled optical waveguides \cite{schwartz_transport_2007, martin_anderson_2011} and cold atoms \cite{roati_anderson_2008, billy_direct_2008, kondov_three-dimensional_2011, jendrzejewski_three-dimensional_2012}.  For such partially disordered systems, the well-accepted scaling theory of Anderson localization does not work, neither the concept of mobility edge. Our results imply that any energy stored in the localized states will not be carried away by energy transport in the eigen-channels of extended states although the energies in these  two types of states overlap both spectrally and spatially. 

% 
%--------------------------------------------------------------------------------------------------------------------------
%--------------------------------------------------------------------------------------------------------------------------
%--------------------------------------------------------------------------------------------------------------------------
%--------------------------------------------------------------------------------------------------------------------------
%--------------------------------------------------------------------------------------------------------------------------

\acknowledgements{This work is supported by Research Grants Council Hong Kong (AoE/P-02/12).}\\
*Corresponding author: {\color{blue}phchan@ust.hk}

%\bibliographystyle{apsrev4-1}
%\bibliography{Coexistence_references}

\begin{thebibliography}{52}%
\makeatletter
\providecommand \@ifxundefined [1]{%
 \@ifx{#1\undefined}
}%
\providecommand \@ifnum [1]{%
 \ifnum #1\expandafter \@firstoftwo
 \else \expandafter \@secondoftwo
 \fi
}%
\providecommand \@ifx [1]{%
 \ifx #1\expandafter \@firstoftwo
 \else \expandafter \@secondoftwo
 \fi
}%
\providecommand \natexlab [1]{#1}%
\providecommand \enquote  [1]{``#1''}%
\providecommand \bibnamefont  [1]{#1}%
\providecommand \bibfnamefont [1]{#1}%
\providecommand \citenamefont [1]{#1}%
\providecommand \href@noop [0]{\@secondoftwo}%
\providecommand \href [0]{\begingroup \@sanitize@url \@href}%
\providecommand \@href[1]{\@@startlink{#1}\@@href}%
\providecommand \@@href[1]{\endgroup#1\@@endlink}%
\providecommand \@sanitize@url [0]{\catcode `\\12\catcode `\$12\catcode
  `\&12\catcode `\#12\catcode `\^12\catcode `\_12\catcode `\%12\relax}%
\providecommand \@@startlink[1]{}%
\providecommand \@@endlink[0]{}%
\providecommand \url  [0]{\begingroup\@sanitize@url \@url }%
\providecommand \@url [1]{\endgroup\@href {#1}{\urlprefix }}%
\providecommand \urlprefix  [0]{URL }%
\providecommand \Eprint [0]{\href }%
\providecommand \doibase [0]{http://dx.doi.org/}%
\providecommand \selectlanguage [0]{\@gobble}%
\providecommand \bibinfo  [0]{\@secondoftwo}%
\providecommand \bibfield  [0]{\@secondoftwo}%
\providecommand \translation [1]{[#1]}%
\providecommand \BibitemOpen [0]{}%
\providecommand \bibitemStop [0]{}%
\providecommand \bibitemNoStop [0]{.\EOS\space}%
\providecommand \EOS [0]{\spacefactor3000\relax}%
\providecommand \BibitemShut  [1]{\csname bibitem#1\endcsname}%
\let\auto@bib@innerbib\@empty
%</preamble>
\bibitem [{\citenamefont {Sheng}(2006)}]{sheng_introduction_2006}%
  \BibitemOpen
  \bibfield  {author} {\bibinfo {author} {\bibfnamefont {P.}~\bibnamefont
  {Sheng}},\ }\href@noop {} {\emph {\bibinfo {title} {Introduction to {Wave}
  {Scattering}, {Localization} and {Mesoscopic} {Phenomena}}}},\ \bibinfo
  {edition} {2nd}\ ed.\ (\bibinfo  {publisher} {Springer-Verlag Berlin
  Heidelberg},\ \bibinfo {year} {2006})\BibitemShut {NoStop}%
\bibitem [{\citenamefont {Lagendijk}\ \emph {et~al.}(2009)\citenamefont
  {Lagendijk}, \citenamefont {Tiggelen},\ and\ \citenamefont
  {Wiersma}}]{lagendijk_fifty_2009}%
  \BibitemOpen
  \bibfield  {author} {\bibinfo {author} {\bibfnamefont {A.}~\bibnamefont
  {Lagendijk}}, \bibinfo {author} {\bibfnamefont {B.~v.}\ \bibnamefont
  {Tiggelen}}, \ and\ \bibinfo {author} {\bibfnamefont {D.~S.}\ \bibnamefont
  {Wiersma}},\ }\href {\doibase 10.1063/1.3206091} {\bibfield  {journal}
  {\bibinfo  {journal} {Physics Today}\ }\textbf {\bibinfo {volume} {62}},\
  \bibinfo {pages} {24} (\bibinfo {year} {2009})}\BibitemShut {NoStop}%
\bibitem [{\citenamefont {Abrahams}(2010)}]{abrahams_50_2010}%
  \BibitemOpen
  \bibfield  {author} {\bibinfo {author} {\bibfnamefont {E.}~\bibnamefont
  {Abrahams}},\ }\href@noop {} {\emph {\bibinfo {title} {50 {Years} of
  {Anderson} {Localization}}}}\ (\bibinfo  {publisher} {world scientific},\
  \bibinfo {year} {2010})\BibitemShut {NoStop}%
\bibitem [{\citenamefont {Anderson}(1958)}]{anderson_absence_1958}%
  \BibitemOpen
  \bibfield  {author} {\bibinfo {author} {\bibfnamefont {P.~W.}\ \bibnamefont
  {Anderson}},\ }\href {\doibase 10.1103/PhysRev.109.1492} {\bibfield
  {journal} {\bibinfo  {journal} {Phys. Rev.}\ }\textbf {\bibinfo {volume}
  {109}},\ \bibinfo {pages} {1492} (\bibinfo {year} {1958})}\BibitemShut
  {NoStop}%
\bibitem [{\citenamefont {De~Raedt}\ \emph {et~al.}(1989)\citenamefont
  {De~Raedt}, \citenamefont {Lagendijk},\ and\ \citenamefont
  {de~Vries}}]{de_raedt_transverse_1989}%
  \BibitemOpen
  \bibfield  {author} {\bibinfo {author} {\bibfnamefont {H.}~\bibnamefont
  {De~Raedt}}, \bibinfo {author} {\bibfnamefont {A.}~\bibnamefont {Lagendijk}},
  \ and\ \bibinfo {author} {\bibfnamefont {P.}~\bibnamefont {de~Vries}},\
  }\href {\doibase 10.1103/PhysRevLett.62.47} {\bibfield  {journal} {\bibinfo
  {journal} {Phys. Rev. Lett.}\ }\textbf {\bibinfo {volume} {62}},\ \bibinfo
  {pages} {47} (\bibinfo {year} {1989})}\BibitemShut {NoStop}%
\bibitem [{\citenamefont {Dunlap}\ \emph {et~al.}(1990)\citenamefont {Dunlap},
  \citenamefont {Wu},\ and\ \citenamefont {Phillips}}]{dunlap_absence_1990}%
  \BibitemOpen
  \bibfield  {author} {\bibinfo {author} {\bibfnamefont {D.~H.}\ \bibnamefont
  {Dunlap}}, \bibinfo {author} {\bibfnamefont {H.-L.}\ \bibnamefont {Wu}}, \
  and\ \bibinfo {author} {\bibfnamefont {P.~W.}\ \bibnamefont {Phillips}},\
  }\href {\doibase 10.1103/PhysRevLett.65.88} {\bibfield  {journal} {\bibinfo
  {journal} {Phys. Rev. Lett.}\ }\textbf {\bibinfo {volume} {65}},\ \bibinfo
  {pages} {88} (\bibinfo {year} {1990})}\BibitemShut {NoStop}%
\bibitem [{\citenamefont {Ye}\ \emph {et~al.}(1992)\citenamefont {Ye},
  \citenamefont {Cody}, \citenamefont {Zhou}, \citenamefont {Sheng},\ and\
  \citenamefont {Norris}}]{ye_observation_1992}%
  \BibitemOpen
  \bibfield  {author} {\bibinfo {author} {\bibfnamefont {L.}~\bibnamefont
  {Ye}}, \bibinfo {author} {\bibfnamefont {G.}~\bibnamefont {Cody}}, \bibinfo
  {author} {\bibfnamefont {M.}~\bibnamefont {Zhou}}, \bibinfo {author}
  {\bibfnamefont {P.}~\bibnamefont {Sheng}}, \ and\ \bibinfo {author}
  {\bibfnamefont {A.~N.}\ \bibnamefont {Norris}},\ }\href {\doibase
  10.1103/PhysRevLett.69.3080} {\bibfield  {journal} {\bibinfo  {journal}
  {Phys. Rev. Lett.}\ }\textbf {\bibinfo {volume} {69}},\ \bibinfo {pages}
  {3080} (\bibinfo {year} {1992})}\BibitemShut {NoStop}%
\bibitem [{\citenamefont {Wiersma}\ \emph {et~al.}(1997)\citenamefont
  {Wiersma}, \citenamefont {Bartolini}, \citenamefont {Lagendijk},\ and\
  \citenamefont {Righini}}]{wiersma_localization_1997}%
  \BibitemOpen
  \bibfield  {author} {\bibinfo {author} {\bibfnamefont {D.~S.}\ \bibnamefont
  {Wiersma}}, \bibinfo {author} {\bibfnamefont {P.}~\bibnamefont {Bartolini}},
  \bibinfo {author} {\bibfnamefont {A.}~\bibnamefont {Lagendijk}}, \ and\
  \bibinfo {author} {\bibfnamefont {R.}~\bibnamefont {Righini}},\ }\href
  {\doibase 10.1038/37757} {\bibfield  {journal} {\bibinfo  {journal} {Nature}\
  }\textbf {\bibinfo {volume} {390}},\ \bibinfo {pages} {671} (\bibinfo {year}
  {1997})}\BibitemShut {NoStop}%
\bibitem [{\citenamefont {Chabanov}\ \emph {et~al.}(2000)\citenamefont
  {Chabanov}, \citenamefont {Stoytchev},\ and\ \citenamefont
  {Genack}}]{chabanov_statistical_2000}%
  \BibitemOpen
  \bibfield  {author} {\bibinfo {author} {\bibfnamefont {A.~A.}\ \bibnamefont
  {Chabanov}}, \bibinfo {author} {\bibfnamefont {M.}~\bibnamefont {Stoytchev}},
  \ and\ \bibinfo {author} {\bibfnamefont {A.~Z.}\ \bibnamefont {Genack}},\
  }\href {\doibase 10.1038/35009055} {\bibfield  {journal} {\bibinfo  {journal}
  {Nature}\ }\textbf {\bibinfo {volume} {404}},\ \bibinfo {pages} {850}
  (\bibinfo {year} {2000})}\BibitemShut {NoStop}%
\bibitem [{\citenamefont {St\"{o}rzer}\ \emph {et~al.}(2006)\citenamefont
  {St\"{o}rzer}, \citenamefont {Gross}, \citenamefont {Aegerter},\ and\
  \citenamefont {Maret}}]{storzer_observation_2006}%
  \BibitemOpen
  \bibfield  {author} {\bibinfo {author} {\bibfnamefont {M.}~\bibnamefont
  {St\"{o}rzer}}, \bibinfo {author} {\bibfnamefont {P.}~\bibnamefont {Gross}},
  \bibinfo {author} {\bibfnamefont {C.~M.}\ \bibnamefont {Aegerter}}, \ and\
  \bibinfo {author} {\bibfnamefont {G.}~\bibnamefont {Maret}},\ }\href
  {\doibase 10.1103/PhysRevLett.96.063904} {\bibfield  {journal} {\bibinfo
  {journal} {Phys. Rev. Lett.}\ }\textbf {\bibinfo {volume} {96}},\ \bibinfo
  {pages} {063904} (\bibinfo {year} {2006})}\BibitemShut {NoStop}%
\bibitem [{\citenamefont {Topolancik}\ \emph {et~al.}(2007)\citenamefont
  {Topolancik}, \citenamefont {Ilic},\ and\ \citenamefont
  {Vollmer}}]{topolancik_experimental_2007}%
  \BibitemOpen
  \bibfield  {author} {\bibinfo {author} {\bibfnamefont {J.}~\bibnamefont
  {Topolancik}}, \bibinfo {author} {\bibfnamefont {B.}~\bibnamefont {Ilic}}, \
  and\ \bibinfo {author} {\bibfnamefont {F.}~\bibnamefont {Vollmer}},\ }\href
  {\doibase 10.1103/PhysRevLett.99.253901} {\bibfield  {journal} {\bibinfo
  {journal} {Phys. Rev. Lett.}\ }\textbf {\bibinfo {volume} {99}},\ \bibinfo
  {pages} {253901} (\bibinfo {year} {2007})}\BibitemShut {NoStop}%
\bibitem [{\citenamefont {Schwartz}\ \emph {et~al.}(2007)\citenamefont
  {Schwartz}, \citenamefont {Bartal}, \citenamefont {Fishman},\ and\
  \citenamefont {Segev}}]{schwartz_transport_2007}%
  \BibitemOpen
  \bibfield  {author} {\bibinfo {author} {\bibfnamefont {T.}~\bibnamefont
  {Schwartz}}, \bibinfo {author} {\bibfnamefont {G.}~\bibnamefont {Bartal}},
  \bibinfo {author} {\bibfnamefont {S.}~\bibnamefont {Fishman}}, \ and\
  \bibinfo {author} {\bibfnamefont {M.}~\bibnamefont {Segev}},\ }\href
  {\doibase 10.1038/nature05623} {\bibfield  {journal} {\bibinfo  {journal}
  {Nature}\ }\textbf {\bibinfo {volume} {446}},\ \bibinfo {pages} {52}
  (\bibinfo {year} {2007})}\BibitemShut {NoStop}%
\bibitem [{\citenamefont {Roati}\ \emph {et~al.}(2008)\citenamefont {Roati},
  \citenamefont {D'Errico}, \citenamefont {Fallani}, \citenamefont {Fattori},
  \citenamefont {Fort}, \citenamefont {Zaccanti}, \citenamefont {Modugno},
  \citenamefont {Modugno},\ and\ \citenamefont
  {Inguscio}}]{roati_anderson_2008}%
  \BibitemOpen
  \bibfield  {author} {\bibinfo {author} {\bibfnamefont {G.}~\bibnamefont
  {Roati}}, \bibinfo {author} {\bibfnamefont {C.}~\bibnamefont {D'Errico}},
  \bibinfo {author} {\bibfnamefont {L.}~\bibnamefont {Fallani}}, \bibinfo
  {author} {\bibfnamefont {M.}~\bibnamefont {Fattori}}, \bibinfo {author}
  {\bibfnamefont {C.}~\bibnamefont {Fort}}, \bibinfo {author} {\bibfnamefont
  {M.}~\bibnamefont {Zaccanti}}, \bibinfo {author} {\bibfnamefont
  {G.}~\bibnamefont {Modugno}}, \bibinfo {author} {\bibfnamefont
  {M.}~\bibnamefont {Modugno}}, \ and\ \bibinfo {author} {\bibfnamefont
  {M.}~\bibnamefont {Inguscio}},\ }\href {\doibase 10.1038/nature07071}
  {\bibfield  {journal} {\bibinfo  {journal} {Nature}\ }\textbf {\bibinfo
  {volume} {453}},\ \bibinfo {pages} {895} (\bibinfo {year}
  {2008})}\BibitemShut {NoStop}%
\bibitem [{\citenamefont {Evers}\ and\ \citenamefont
  {Mirlin}(2008)}]{evers_anderson_2008}%
  \BibitemOpen
  \bibfield  {author} {\bibinfo {author} {\bibfnamefont {F.}~\bibnamefont
  {Evers}}\ and\ \bibinfo {author} {\bibfnamefont {A.~D.}\ \bibnamefont
  {Mirlin}},\ }\href {\doibase 10.1103/RevModPhys.80.1355} {\bibfield
  {journal} {\bibinfo  {journal} {Rev. Mod. Phys.}\ }\textbf {\bibinfo {volume}
  {80}},\ \bibinfo {pages} {1355} (\bibinfo {year} {2008})}\BibitemShut
  {NoStop}%
\bibitem [{\citenamefont {Lahini}\ \emph {et~al.}(2008)\citenamefont {Lahini},
  \citenamefont {Avidan}, \citenamefont {Pozzi}, \citenamefont {Sorel},
  \citenamefont {Morandotti}, \citenamefont {Christodoulides},\ and\
  \citenamefont {Silberberg}}]{lahini_anderson_2008}%
  \BibitemOpen
  \bibfield  {author} {\bibinfo {author} {\bibfnamefont {Y.}~\bibnamefont
  {Lahini}}, \bibinfo {author} {\bibfnamefont {A.}~\bibnamefont {Avidan}},
  \bibinfo {author} {\bibfnamefont {F.}~\bibnamefont {Pozzi}}, \bibinfo
  {author} {\bibfnamefont {M.}~\bibnamefont {Sorel}}, \bibinfo {author}
  {\bibfnamefont {R.}~\bibnamefont {Morandotti}}, \bibinfo {author}
  {\bibfnamefont {D.~N.}\ \bibnamefont {Christodoulides}}, \ and\ \bibinfo
  {author} {\bibfnamefont {Y.}~\bibnamefont {Silberberg}},\ }\href {\doibase
  10.1103/PhysRevLett.100.013906} {\bibfield  {journal} {\bibinfo  {journal}
  {Phys. Rev. Lett.}\ }\textbf {\bibinfo {volume} {100}},\ \bibinfo {pages}
  {013906} (\bibinfo {year} {2008})}\BibitemShut {NoStop}%
\bibitem [{\citenamefont {Hu}\ \emph {et~al.}(2008)\citenamefont {Hu},
  \citenamefont {Strybulevych}, \citenamefont {Page}, \citenamefont
  {Skipetrov},\ and\ \citenamefont {van Tiggelen}}]{hu_localization_2008}%
  \BibitemOpen
  \bibfield  {author} {\bibinfo {author} {\bibfnamefont {H.}~\bibnamefont
  {Hu}}, \bibinfo {author} {\bibfnamefont {A.}~\bibnamefont {Strybulevych}},
  \bibinfo {author} {\bibfnamefont {J.~H.}\ \bibnamefont {Page}}, \bibinfo
  {author} {\bibfnamefont {S.~E.}\ \bibnamefont {Skipetrov}}, \ and\ \bibinfo
  {author} {\bibfnamefont {B.~A.}\ \bibnamefont {van Tiggelen}},\ }\href
  {\doibase 10.1038/nphys1101} {\bibfield  {journal} {\bibinfo  {journal} {Nat
  Phys}\ }\textbf {\bibinfo {volume} {4}},\ \bibinfo {pages} {945} (\bibinfo
  {year} {2008})}\BibitemShut {NoStop}%
\bibitem [{\citenamefont {Chab\'{e}}\ \emph {et~al.}(2008)\citenamefont
  {Chab\'{e}}, \citenamefont {Lemari\'{e}}, \citenamefont {Gr\'{e}maud},
  \citenamefont {Delande}, \citenamefont {Szriftgiser},\ and\ \citenamefont
  {Garreau}}]{chabe_experimental_2008}%
  \BibitemOpen
  \bibfield  {author} {\bibinfo {author} {\bibfnamefont {J.}~\bibnamefont
  {Chab\'{e}}}, \bibinfo {author} {\bibfnamefont {G.}~\bibnamefont
  {Lemari\'{e}}}, \bibinfo {author} {\bibfnamefont {B.}~\bibnamefont
  {Gr\'{e}maud}}, \bibinfo {author} {\bibfnamefont {D.}~\bibnamefont
  {Delande}}, \bibinfo {author} {\bibfnamefont {P.}~\bibnamefont
  {Szriftgiser}}, \ and\ \bibinfo {author} {\bibfnamefont {J.~C.}\ \bibnamefont
  {Garreau}},\ }\href {\doibase 10.1103/PhysRevLett.101.255702} {\bibfield
  {journal} {\bibinfo  {journal} {Phys. Rev. Lett.}\ }\textbf {\bibinfo
  {volume} {101}},\ \bibinfo {pages} {255702} (\bibinfo {year}
  {2008})}\BibitemShut {NoStop}%
\bibitem [{\citenamefont {Billy}\ \emph {et~al.}(2008)\citenamefont {Billy},
  \citenamefont {Josse}, \citenamefont {Zuo}, \citenamefont {Bernard},
  \citenamefont {Hambrecht}, \citenamefont {Lugan}, \citenamefont
  {Cl\'{e}ment}, \citenamefont {Sanchez-Palencia}, \citenamefont {Bouyer},\
  and\ \citenamefont {Aspect}}]{billy_direct_2008}%
  \BibitemOpen
  \bibfield  {author} {\bibinfo {author} {\bibfnamefont {J.}~\bibnamefont
  {Billy}}, \bibinfo {author} {\bibfnamefont {V.}~\bibnamefont {Josse}},
  \bibinfo {author} {\bibfnamefont {Z.}~\bibnamefont {Zuo}}, \bibinfo {author}
  {\bibfnamefont {A.}~\bibnamefont {Bernard}}, \bibinfo {author} {\bibfnamefont
  {B.}~\bibnamefont {Hambrecht}}, \bibinfo {author} {\bibfnamefont
  {P.}~\bibnamefont {Lugan}}, \bibinfo {author} {\bibfnamefont
  {D.}~\bibnamefont {Cl\'{e}ment}}, \bibinfo {author} {\bibfnamefont
  {L.}~\bibnamefont {Sanchez-Palencia}}, \bibinfo {author} {\bibfnamefont
  {P.}~\bibnamefont {Bouyer}}, \ and\ \bibinfo {author} {\bibfnamefont
  {A.}~\bibnamefont {Aspect}},\ }\href {\doibase 10.1038/nature07000}
  {\bibfield  {journal} {\bibinfo  {journal} {Nature}\ }\textbf {\bibinfo
  {volume} {453}},\ \bibinfo {pages} {891} (\bibinfo {year}
  {2008})}\BibitemShut {NoStop}%
\bibitem [{\citenamefont {Lahini}\ \emph {et~al.}(2009)\citenamefont {Lahini},
  \citenamefont {Pugatch}, \citenamefont {Pozzi}, \citenamefont {Sorel},
  \citenamefont {Morandotti}, \citenamefont {Davidson},\ and\ \citenamefont
  {Silberberg}}]{lahini_observation_2009}%
  \BibitemOpen
  \bibfield  {author} {\bibinfo {author} {\bibfnamefont {Y.}~\bibnamefont
  {Lahini}}, \bibinfo {author} {\bibfnamefont {R.}~\bibnamefont {Pugatch}},
  \bibinfo {author} {\bibfnamefont {F.}~\bibnamefont {Pozzi}}, \bibinfo
  {author} {\bibfnamefont {M.}~\bibnamefont {Sorel}}, \bibinfo {author}
  {\bibfnamefont {R.}~\bibnamefont {Morandotti}}, \bibinfo {author}
  {\bibfnamefont {N.}~\bibnamefont {Davidson}}, \ and\ \bibinfo {author}
  {\bibfnamefont {Y.}~\bibnamefont {Silberberg}},\ }\href {\doibase
  10.1103/PhysRevLett.103.013901} {\bibfield  {journal} {\bibinfo  {journal}
  {Phys. Rev. Lett.}\ }\textbf {\bibinfo {volume} {103}},\ \bibinfo {pages}
  {013901} (\bibinfo {year} {2009})}\BibitemShut {NoStop}%
\bibitem [{\citenamefont {Kondov}\ \emph {et~al.}(2011)\citenamefont {Kondov},
  \citenamefont {McGehee}, \citenamefont {Zirbel},\ and\ \citenamefont
  {DeMarco}}]{kondov_three-dimensional_2011}%
  \BibitemOpen
  \bibfield  {author} {\bibinfo {author} {\bibfnamefont {S.~S.}\ \bibnamefont
  {Kondov}}, \bibinfo {author} {\bibfnamefont {W.~R.}\ \bibnamefont {McGehee}},
  \bibinfo {author} {\bibfnamefont {J.~J.}\ \bibnamefont {Zirbel}}, \ and\
  \bibinfo {author} {\bibfnamefont {B.}~\bibnamefont {DeMarco}},\ }\href
  {\doibase 10.1126/science.1209019} {\bibfield  {journal} {\bibinfo  {journal}
  {Science}\ }\textbf {\bibinfo {volume} {334}},\ \bibinfo {pages} {66}
  (\bibinfo {year} {2011})}\BibitemShut {NoStop}%
\bibitem [{\citenamefont {Jendrzejewski}\ \emph {et~al.}(2012)\citenamefont
  {Jendrzejewski}, \citenamefont {Bernard}, \citenamefont {M\"{u}ller},
  \citenamefont {Cheinet}, \citenamefont {Josse}, \citenamefont {Piraud},
  \citenamefont {Pezz\'{e}}, \citenamefont {Sanchez-Palencia}, \citenamefont
  {Aspect},\ and\ \citenamefont
  {Bouyer}}]{jendrzejewski_three-dimensional_2012}%
  \BibitemOpen
  \bibfield  {author} {\bibinfo {author} {\bibfnamefont {F.}~\bibnamefont
  {Jendrzejewski}}, \bibinfo {author} {\bibfnamefont {A.}~\bibnamefont
  {Bernard}}, \bibinfo {author} {\bibfnamefont {K.}~\bibnamefont {M\"{u}ller}},
  \bibinfo {author} {\bibfnamefont {P.}~\bibnamefont {Cheinet}}, \bibinfo
  {author} {\bibfnamefont {V.}~\bibnamefont {Josse}}, \bibinfo {author}
  {\bibfnamefont {M.}~\bibnamefont {Piraud}}, \bibinfo {author} {\bibfnamefont
  {L.}~\bibnamefont {Pezz\'{e}}}, \bibinfo {author} {\bibfnamefont
  {L.}~\bibnamefont {Sanchez-Palencia}}, \bibinfo {author} {\bibfnamefont
  {A.}~\bibnamefont {Aspect}}, \ and\ \bibinfo {author} {\bibfnamefont
  {P.}~\bibnamefont {Bouyer}},\ }\href {\doibase 10.1038/nphys2256} {\bibfield
  {journal} {\bibinfo  {journal} {Nat Phys}\ }\textbf {\bibinfo {volume} {8}},\
  \bibinfo {pages} {398} (\bibinfo {year} {2012})}\BibitemShut {NoStop}%
\bibitem [{\citenamefont {Segev}\ \emph {et~al.}(2013)\citenamefont {Segev},
  \citenamefont {Silberberg},\ and\ \citenamefont
  {Christodoulides}}]{segev_anderson_2013}%
  \BibitemOpen
  \bibfield  {author} {\bibinfo {author} {\bibfnamefont {M.}~\bibnamefont
  {Segev}}, \bibinfo {author} {\bibfnamefont {Y.}~\bibnamefont {Silberberg}}, \
  and\ \bibinfo {author} {\bibfnamefont {D.~N.}\ \bibnamefont
  {Christodoulides}},\ }\href {\doibase 10.1038/nphoton.2013.30} {\bibfield
  {journal} {\bibinfo  {journal} {Nat Photon}\ }\textbf {\bibinfo {volume}
  {7}},\ \bibinfo {pages} {197} (\bibinfo {year} {2013})}\BibitemShut {NoStop}%
\bibitem [{\citenamefont {Hsieh}\ \emph {et~al.}(2015)\citenamefont {Hsieh},
  \citenamefont {Chung}, \citenamefont {McMillan}, \citenamefont {Tsai},
  \citenamefont {Lu}, \citenamefont {Panoiu},\ and\ \citenamefont
  {Wong}}]{hsieh_photon_2015}%
  \BibitemOpen
  \bibfield  {author} {\bibinfo {author} {\bibfnamefont {P.}~\bibnamefont
  {Hsieh}}, \bibinfo {author} {\bibfnamefont {C.}~\bibnamefont {Chung}},
  \bibinfo {author} {\bibfnamefont {J.~F.}\ \bibnamefont {McMillan}}, \bibinfo
  {author} {\bibfnamefont {M.}~\bibnamefont {Tsai}}, \bibinfo {author}
  {\bibfnamefont {M.}~\bibnamefont {Lu}}, \bibinfo {author} {\bibfnamefont
  {N.~C.}\ \bibnamefont {Panoiu}}, \ and\ \bibinfo {author} {\bibfnamefont
  {C.~W.}\ \bibnamefont {Wong}},\ }\href {\doibase 10.1038/nphys3211}
  {\bibfield  {journal} {\bibinfo  {journal} {Nat Phys}\ }\textbf {\bibinfo
  {volume} {11}},\ \bibinfo {pages} {268} (\bibinfo {year} {2015})}\BibitemShut
  {NoStop}%
\bibitem [{\citenamefont {Abrahams}\ \emph {et~al.}(1979)\citenamefont
  {Abrahams}, \citenamefont {Anderson}, \citenamefont {Licciardello},\ and\
  \citenamefont {Ramakrishnan}}]{abrahams_scaling_1979}%
  \BibitemOpen
  \bibfield  {author} {\bibinfo {author} {\bibfnamefont {E.}~\bibnamefont
  {Abrahams}}, \bibinfo {author} {\bibfnamefont {P.~W.}\ \bibnamefont
  {Anderson}}, \bibinfo {author} {\bibfnamefont {D.~C.}\ \bibnamefont
  {Licciardello}}, \ and\ \bibinfo {author} {\bibfnamefont {T.~V.}\
  \bibnamefont {Ramakrishnan}},\ }\href {\doibase 10.1103/PhysRevLett.42.673}
  {\bibfield  {journal} {\bibinfo  {journal} {Phys. Rev. Lett.}\ }\textbf
  {\bibinfo {volume} {42}},\ \bibinfo {pages} {673} (\bibinfo {year}
  {1979})}\BibitemShut {NoStop}%
\bibitem [{\citenamefont {Mott}(1990)}]{mott_metal-insulator_1990}%
  \BibitemOpen
  \bibfield  {author} {\bibinfo {author} {\bibfnamefont {N.~F.}\ \bibnamefont
  {Mott}},\ }\href@noop {} {\emph {\bibinfo {title} {Metal-insulator
  transitions}}}\ (\bibinfo  {publisher} {Taylor \& Francis},\ \bibinfo
  {address} {London},\ \bibinfo {year} {1990})\BibitemShut {NoStop}%
\bibitem [{\citenamefont {Economou}\ and\ \citenamefont
  {Cohen}(1970)}]{economou_localization_1970}%
  \BibitemOpen
  \bibfield  {author} {\bibinfo {author} {\bibfnamefont {E.~N.}\ \bibnamefont
  {Economou}}\ and\ \bibinfo {author} {\bibfnamefont {M.~H.}\ \bibnamefont
  {Cohen}},\ }\href
  {http://journals.aps.org/prl/abstract/10.1103/PhysRevLett.25.1445} {\bibfield
   {journal} {\bibinfo  {journal} {Phys. Rev. Lett.}\ }\textbf {\bibinfo
  {volume} {25}},\ \bibinfo {pages} {1445} (\bibinfo {year}
  {1970})}\BibitemShut {NoStop}%
\bibitem [{\citenamefont {Licciardello}\ and\ \citenamefont
  {Thouless}(1975)}]{licciardello_conductivity_1975}%
  \BibitemOpen
  \bibfield  {author} {\bibinfo {author} {\bibfnamefont {D.~C.}\ \bibnamefont
  {Licciardello}}\ and\ \bibinfo {author} {\bibfnamefont {D.~J.}\ \bibnamefont
  {Thouless}},\ }\href {\doibase 10.1088/0022-3719/8/24/009} {\bibfield
  {journal} {\bibinfo  {journal} {J. Phys. C: Solid State Phys.}\ }\textbf
  {\bibinfo {volume} {8}},\ \bibinfo {pages} {4157} (\bibinfo {year}
  {1975})}\BibitemShut {NoStop}%
\bibitem [{\citenamefont {John}(1984)}]{john_electromagnetic_1984}%
  \BibitemOpen
  \bibfield  {author} {\bibinfo {author} {\bibfnamefont {S.}~\bibnamefont
  {John}},\ }\href {\doibase 10.1103/PhysRevLett.53.2169} {\bibfield  {journal}
  {\bibinfo  {journal} {Phys. Rev. Lett.}\ }\textbf {\bibinfo {volume} {53}},\
  \bibinfo {pages} {2169} (\bibinfo {year} {1984})}\BibitemShut {NoStop}%
\bibitem [{\citenamefont {Kuhl}\ \emph {et~al.}(2000)\citenamefont {Kuhl},
  \citenamefont {Izrailev}, \citenamefont {Krokhin},\ and\ \citenamefont
  {St\"{o}ckmann}}]{kuhl_experimental_2000}%
  \BibitemOpen
  \bibfield  {author} {\bibinfo {author} {\bibfnamefont {U.}~\bibnamefont
  {Kuhl}}, \bibinfo {author} {\bibfnamefont {F.~M.}\ \bibnamefont {Izrailev}},
  \bibinfo {author} {\bibfnamefont {A.~A.}\ \bibnamefont {Krokhin}}, \ and\
  \bibinfo {author} {\bibfnamefont {H.-J.}\ \bibnamefont {St\"{o}ckmann}},\
  }\href {\doibase 10.1063/1.127068} {\bibfield  {journal} {\bibinfo  {journal}
  {Appl. Phys. Lett.}\ }\textbf {\bibinfo {volume} {77}},\ \bibinfo {pages}
  {633} (\bibinfo {year} {2000})}\BibitemShut {NoStop}%
\bibitem [{\citenamefont {Delande}\ and\ \citenamefont
  {Orso}(2014)}]{delande_mobility_2014}%
  \BibitemOpen
  \bibfield  {author} {\bibinfo {author} {\bibfnamefont {D.}~\bibnamefont
  {Delande}}\ and\ \bibinfo {author} {\bibfnamefont {G.}~\bibnamefont {Orso}},\
  }\href {\doibase 10.1103/PhysRevLett.113.060601} {\bibfield  {journal}
  {\bibinfo  {journal} {Phys. Rev. Lett.}\ }\textbf {\bibinfo {volume} {113}},\
  \bibinfo {pages} {060601} (\bibinfo {year} {2014})}\BibitemShut {NoStop}%
\bibitem [{\citenamefont {Semeghini}\ \emph {et~al.}(2015)\citenamefont
  {Semeghini}, \citenamefont {Landini}, \citenamefont {Castilho}, \citenamefont
  {Roy}, \citenamefont {Spagnolli}, \citenamefont {Trenkwalder}, \citenamefont
  {Fattori}, \citenamefont {Inguscio},\ and\ \citenamefont
  {Modugno}}]{semeghini_measurement_2015}%
  \BibitemOpen
  \bibfield  {author} {\bibinfo {author} {\bibfnamefont {G.}~\bibnamefont
  {Semeghini}}, \bibinfo {author} {\bibfnamefont {M.}~\bibnamefont {Landini}},
  \bibinfo {author} {\bibfnamefont {P.}~\bibnamefont {Castilho}}, \bibinfo
  {author} {\bibfnamefont {S.}~\bibnamefont {Roy}}, \bibinfo {author}
  {\bibfnamefont {G.}~\bibnamefont {Spagnolli}}, \bibinfo {author}
  {\bibfnamefont {A.}~\bibnamefont {Trenkwalder}}, \bibinfo {author}
  {\bibfnamefont {M.}~\bibnamefont {Fattori}}, \bibinfo {author} {\bibfnamefont
  {M.}~\bibnamefont {Inguscio}}, \ and\ \bibinfo {author} {\bibfnamefont
  {G.}~\bibnamefont {Modugno}},\ }\href {\doibase 10.1038/nphys3339} {\bibfield
   {journal} {\bibinfo  {journal} {Nat Phys}\ }\textbf {\bibinfo {volume}
  {11}},\ \bibinfo {pages} {554} (\bibinfo {year} {2015})}\BibitemShut
  {NoStop}%
\bibitem [{\citenamefont {Sheng}\ and\ \citenamefont
  {Zhang}(1986)}]{sheng_scalar-wave_1986}%
  \BibitemOpen
  \bibfield  {author} {\bibinfo {author} {\bibfnamefont {P.}~\bibnamefont
  {Sheng}}\ and\ \bibinfo {author} {\bibfnamefont {Z.-Q.}\ \bibnamefont
  {Zhang}},\ }\href {\doibase 10.1103/PhysRevLett.57.1879} {\bibfield
  {journal} {\bibinfo  {journal} {Phys. Rev. Lett.}\ }\textbf {\bibinfo
  {volume} {57}},\ \bibinfo {pages} {1879} (\bibinfo {year}
  {1986})}\BibitemShut {NoStop}%
\bibitem [{\citenamefont {Li}\ \emph {et~al.}(1989)\citenamefont {Li},
  \citenamefont {Soukoulis}, \citenamefont {Economou},\ and\ \citenamefont
  {Grest}}]{li_anisotropic_1989}%
  \BibitemOpen
  \bibfield  {author} {\bibinfo {author} {\bibfnamefont {Q.}~\bibnamefont
  {Li}}, \bibinfo {author} {\bibfnamefont {C.~M.}\ \bibnamefont {Soukoulis}},
  \bibinfo {author} {\bibfnamefont {E.~N.}\ \bibnamefont {Economou}}, \ and\
  \bibinfo {author} {\bibfnamefont {G.~S.}\ \bibnamefont {Grest}},\ }\href
  {\doibase 10.1103/PhysRevB.40.2825} {\bibfield  {journal} {\bibinfo
  {journal} {Phys. Rev. B}\ }\textbf {\bibinfo {volume} {40}},\ \bibinfo
  {pages} {2825} (\bibinfo {year} {1989})}\BibitemShut {NoStop}%
\bibitem [{\citenamefont {Vollhardt}\ and\ \citenamefont
  {W\"{o}lfle}(1982)}]{vollhardt_scaling_1982}%
  \BibitemOpen
  \bibfield  {author} {\bibinfo {author} {\bibfnamefont {D.}~\bibnamefont
  {Vollhardt}}\ and\ \bibinfo {author} {\bibfnamefont {P.}~\bibnamefont
  {W\"{o}lfle}},\ }\href {\doibase 10.1103/PhysRevLett.48.699} {\bibfield
  {journal} {\bibinfo  {journal} {Phys. Rev. Lett.}\ }\textbf {\bibinfo
  {volume} {48}},\ \bibinfo {pages} {699} (\bibinfo {year} {1982})}\BibitemShut
  {NoStop}%
\bibitem [{\citenamefont {Biddle}\ and\ \citenamefont
  {Das~Sarma}(2010)}]{biddle_predicted_2010}%
  \BibitemOpen
  \bibfield  {author} {\bibinfo {author} {\bibfnamefont {J.}~\bibnamefont
  {Biddle}}\ and\ \bibinfo {author} {\bibfnamefont {S.}~\bibnamefont
  {Das~Sarma}},\ }\href {\doibase 10.1103/PhysRevLett.104.070601} {\bibfield
  {journal} {\bibinfo  {journal} {Phys. Rev. Lett.}\ }\textbf {\bibinfo
  {volume} {104}},\ \bibinfo {pages} {070601} (\bibinfo {year}
  {2010})}\BibitemShut {NoStop}%
\bibitem [{\citenamefont {Ganeshan}\ \emph {et~al.}(2015)\citenamefont
  {Ganeshan}, \citenamefont {Pixley},\ and\ \citenamefont
  {Das~Sarma}}]{ganeshan_nearest_2015}%
  \BibitemOpen
  \bibfield  {author} {\bibinfo {author} {\bibfnamefont {S.}~\bibnamefont
  {Ganeshan}}, \bibinfo {author} {\bibfnamefont {J.}~\bibnamefont {Pixley}}, \
  and\ \bibinfo {author} {\bibfnamefont {S.}~\bibnamefont {Das~Sarma}},\ }\href
  {\doibase 10.1103/PhysRevLett.114.146601} {\bibfield  {journal} {\bibinfo
  {journal} {Phys. Rev. Lett.}\ }\textbf {\bibinfo {volume} {114}},\ \bibinfo
  {pages} {146601} (\bibinfo {year} {2015})}\BibitemShut {NoStop}%
\bibitem [{\citenamefont {Pasek}\ \emph {et~al.}(2017)\citenamefont {Pasek},
  \citenamefont {Orso},\ and\ \citenamefont {Delande}}]{pasek_anderson_2017}%
  \BibitemOpen
  \bibfield  {author} {\bibinfo {author} {\bibfnamefont {M.}~\bibnamefont
  {Pasek}}, \bibinfo {author} {\bibfnamefont {G.}~\bibnamefont {Orso}}, \ and\
  \bibinfo {author} {\bibfnamefont {D.}~\bibnamefont {Delande}},\ }\href
  {\doibase 10.1103/PhysRevLett.118.170403} {\bibfield  {journal} {\bibinfo
  {journal} {Phys. Rev. Lett.}\ }\textbf {\bibinfo {volume} {118}},\ \bibinfo
  {pages} {170403} (\bibinfo {year} {2017})}\BibitemShut {NoStop}%
\bibitem [{\citenamefont {von Neuman}\ and\ \citenamefont
  {Wigner}(1929)}]{von_neuman_j._1929}%
  \BibitemOpen
  \bibfield  {author} {\bibinfo {author} {\bibfnamefont {J.}~\bibnamefont {von
  Neuman}}\ and\ \bibinfo {author} {\bibfnamefont {E.}~\bibnamefont {Wigner}},\
  }\href@noop {} {\bibfield  {journal} {\bibinfo  {journal} {Phys. Z.}\
  }\textbf {\bibinfo {volume} {30}},\ \bibinfo {pages} {465} (\bibinfo {year}
  {1929})}\BibitemShut {NoStop}%
\bibitem [{\citenamefont {Friedrich}\ and\ \citenamefont
  {Wintgen}(1985)}]{friedrich_interfering_1985}%
  \BibitemOpen
  \bibfield  {author} {\bibinfo {author} {\bibfnamefont {H.}~\bibnamefont
  {Friedrich}}\ and\ \bibinfo {author} {\bibfnamefont {D.}~\bibnamefont
  {Wintgen}},\ }\href {\doibase 10.1103/PhysRevA.32.3231} {\bibfield  {journal}
  {\bibinfo  {journal} {Phys. Rev. A}\ }\textbf {\bibinfo {volume} {32}},\
  \bibinfo {pages} {3231} (\bibinfo {year} {1985})}\BibitemShut {NoStop}%
\bibitem [{\citenamefont {Hsu}\ \emph {et~al.}(2013)\citenamefont {Hsu},
  \citenamefont {Zhen}, \citenamefont {Lee}, \citenamefont {Chua},
  \citenamefont {Johnson}, \citenamefont {Joannopoulos},\ and\ \citenamefont
  {Solja\v{c}i\'{c}}}]{hsu_observation_2013}%
  \BibitemOpen
  \bibfield  {author} {\bibinfo {author} {\bibfnamefont {C.~W.}\ \bibnamefont
  {Hsu}}, \bibinfo {author} {\bibfnamefont {B.}~\bibnamefont {Zhen}}, \bibinfo
  {author} {\bibfnamefont {J.}~\bibnamefont {Lee}}, \bibinfo {author}
  {\bibfnamefont {S.-L.}\ \bibnamefont {Chua}}, \bibinfo {author}
  {\bibfnamefont {S.~G.}\ \bibnamefont {Johnson}}, \bibinfo {author}
  {\bibfnamefont {J.~D.}\ \bibnamefont {Joannopoulos}}, \ and\ \bibinfo
  {author} {\bibfnamefont {M.}~\bibnamefont {Solja\v{c}i\'{c}}},\ }\href
  {\doibase 10.1038/nature12289} {\bibfield  {journal} {\bibinfo  {journal}
  {Nature}\ }\textbf {\bibinfo {volume} {499}},\ \bibinfo {pages} {188}
  (\bibinfo {year} {2013})}\BibitemShut {NoStop}%
\bibitem [{\citenamefont {Plotnik}\ \emph {et~al.}(2011)\citenamefont
  {Plotnik}, \citenamefont {Peleg}, \citenamefont {Dreisow}, \citenamefont
  {Heinrich}, \citenamefont {Nolte}, \citenamefont {Szameit},\ and\
  \citenamefont {Segev}}]{plotnik_experimental_2011}%
  \BibitemOpen
  \bibfield  {author} {\bibinfo {author} {\bibfnamefont {Y.}~\bibnamefont
  {Plotnik}}, \bibinfo {author} {\bibfnamefont {O.}~\bibnamefont {Peleg}},
  \bibinfo {author} {\bibfnamefont {F.}~\bibnamefont {Dreisow}}, \bibinfo
  {author} {\bibfnamefont {M.}~\bibnamefont {Heinrich}}, \bibinfo {author}
  {\bibfnamefont {S.}~\bibnamefont {Nolte}}, \bibinfo {author} {\bibfnamefont
  {A.}~\bibnamefont {Szameit}}, \ and\ \bibinfo {author} {\bibfnamefont
  {M.}~\bibnamefont {Segev}},\ }\href {\doibase 10.1103/PhysRevLett.107.183901}
  {\bibfield  {journal} {\bibinfo  {journal} {Phys. Rev. Lett.}\ }\textbf
  {\bibinfo {volume} {107}},\ \bibinfo {pages} {183901} (\bibinfo {year}
  {2011})}\BibitemShut {NoStop}%
\bibitem [{\citenamefont {Mur-Petit}\ and\ \citenamefont
  {Molina}(2014)}]{mur-petit_chiral_2014}%
  \BibitemOpen
  \bibfield  {author} {\bibinfo {author} {\bibfnamefont {J.}~\bibnamefont
  {Mur-Petit}}\ and\ \bibinfo {author} {\bibfnamefont {R.~A.}\ \bibnamefont
  {Molina}},\ }\href {\doibase 10.1103/PhysRevB.90.035434} {\bibfield
  {journal} {\bibinfo  {journal} {Phys. Rev. B}\ }\textbf {\bibinfo {volume}
  {90}},\ \bibinfo {pages} {035434} (\bibinfo {year} {2014})}\BibitemShut
  {NoStop}%
\bibitem [{\citenamefont {Zhen}\ \emph {et~al.}(2014)\citenamefont {Zhen},
  \citenamefont {Hsu}, \citenamefont {Lu}, \citenamefont {Stone},\ and\
  \citenamefont {Solja\v{c}i\'{c}}}]{zhen_topological_2014}%
  \BibitemOpen
  \bibfield  {author} {\bibinfo {author} {\bibfnamefont {B.}~\bibnamefont
  {Zhen}}, \bibinfo {author} {\bibfnamefont {C.~W.}\ \bibnamefont {Hsu}},
  \bibinfo {author} {\bibfnamefont {L.}~\bibnamefont {Lu}}, \bibinfo {author}
  {\bibfnamefont {A.~D.}\ \bibnamefont {Stone}}, \ and\ \bibinfo {author}
  {\bibfnamefont {M.}~\bibnamefont {Solja\v{c}i\'{c}}},\ }\href {\doibase
  10.1103/PhysRevLett.113.257401} {\bibfield  {journal} {\bibinfo  {journal}
  {Phys. Rev. Lett.}\ }\textbf {\bibinfo {volume} {113}},\ \bibinfo {pages}
  {257401} (\bibinfo {year} {2014})}\BibitemShut {NoStop}%
\bibitem [{\citenamefont {Xiao}\ \emph {et~al.}(2017)\citenamefont {Xiao},
  \citenamefont {Ma}, \citenamefont {Zhang},\ and\ \citenamefont
  {Chan}}]{xiao_topological_2017}%
  \BibitemOpen
  \bibfield  {author} {\bibinfo {author} {\bibfnamefont {Y.-X.}\ \bibnamefont
  {Xiao}}, \bibinfo {author} {\bibfnamefont {G.}~\bibnamefont {Ma}}, \bibinfo
  {author} {\bibfnamefont {Z.-Q.}\ \bibnamefont {Zhang}}, \ and\ \bibinfo
  {author} {\bibfnamefont {C.~T.}\ \bibnamefont {Chan}},\ }\href {\doibase
  10.1103/PhysRevLett.118.166803} {\bibfield  {journal} {\bibinfo  {journal}
  {Phys. Rev. Lett.}\ }\textbf {\bibinfo {volume} {118}},\ \bibinfo {pages}
  {166803} (\bibinfo {year} {2017})}\BibitemShut {NoStop}%
\bibitem [{\citenamefont {Bulgakov}\ and\ \citenamefont
  {Maksimov}(2017)}]{bulgakov_topological_2017}%
  \BibitemOpen
  \bibfield  {author} {\bibinfo {author} {\bibfnamefont {E.~N.}\ \bibnamefont
  {Bulgakov}}\ and\ \bibinfo {author} {\bibfnamefont {D.~N.}\ \bibnamefont
  {Maksimov}},\ }\href {\doibase 10.1103/PhysRevLett.118.267401} {\bibfield
  {journal} {\bibinfo  {journal} {Phys. Rev. Lett.}\ }\textbf {\bibinfo
  {volume} {118}},\ \bibinfo {pages} {267401} (\bibinfo {year}
  {2017})}\BibitemShut {NoStop}%
\bibitem [{\citenamefont {Hsu}\ \emph {et~al.}(2016)\citenamefont {Hsu},
  \citenamefont {Zhen}, \citenamefont {Stone}, \citenamefont {Joannopoulos},\
  and\ \citenamefont {Solja\v{c}i\'{c}}}]{hsu_bound_2016}%
  \BibitemOpen
  \bibfield  {author} {\bibinfo {author} {\bibfnamefont {C.~W.}\ \bibnamefont
  {Hsu}}, \bibinfo {author} {\bibfnamefont {B.}~\bibnamefont {Zhen}}, \bibinfo
  {author} {\bibfnamefont {A.~D.}\ \bibnamefont {Stone}}, \bibinfo {author}
  {\bibfnamefont {J.~D.}\ \bibnamefont {Joannopoulos}}, \ and\ \bibinfo
  {author} {\bibfnamefont {M.}~\bibnamefont {Solja\v{c}i\'{c}}},\ }\href
  {\doibase 10.1038/natrevmats.2016.48} {\bibfield  {journal} {\bibinfo
  {journal} {Nat. Rev. Mater.}\ }\textbf {\bibinfo {volume} {1}},\ \bibinfo
  {pages} {16048} (\bibinfo {year} {2016})}\BibitemShut {NoStop}%
\bibitem [{\citenamefont {Economou}(2006)}]{economou_greens_2006}%
  \BibitemOpen
  \bibfield  {author} {\bibinfo {author} {\bibfnamefont {E.~N.}\ \bibnamefont
  {Economou}},\ }\href@noop {} {\emph {\bibinfo {title} {Green's {Functions} in
  {Quantum} {Physics}}}},\ \bibinfo {edition} {3rd}\ ed.\ (\bibinfo
  {publisher} {Springer-Verlag Berlin Heidelberg},\ \bibinfo {year}
  {2006})\BibitemShut {NoStop}%
\bibitem [{\citenamefont {Pezz\'{e}}\ and\ \citenamefont
  {Sanchez-Palencia}(2011)}]{pezze_localized_2011}%
  \BibitemOpen
  \bibfield  {author} {\bibinfo {author} {\bibfnamefont {L.}~\bibnamefont
  {Pezz\'{e}}}\ and\ \bibinfo {author} {\bibfnamefont {L.}~\bibnamefont
  {Sanchez-Palencia}},\ }\href {\doibase 10.1103/PhysRevLett.106.040601}
  {\bibfield  {journal} {\bibinfo  {journal} {Phys. Rev. Lett.}\ }\textbf
  {\bibinfo {volume} {106}},\ \bibinfo {pages} {040601} (\bibinfo {year}
  {2011})}\BibitemShut {NoStop}%
\bibitem [{\citenamefont {Rodriguez}\ \emph {et~al.}(2012)\citenamefont
  {Rodriguez}, \citenamefont {Chakrabarti},\ and\ \citenamefont
  {Römer}}]{rodriguez_controlled_2012}%
  \BibitemOpen
  \bibfield  {author} {\bibinfo {author} {\bibfnamefont {A.}~\bibnamefont
  {Rodriguez}}, \bibinfo {author} {\bibfnamefont {A.}~\bibnamefont
  {Chakrabarti}}, \ and\ \bibinfo {author} {\bibfnamefont {R.~A.}\ \bibnamefont
  {Römer}},\ }\href {\doibase 10.1103/PhysRevB.86.085119} {\bibfield
  {journal} {\bibinfo  {journal} {Phys. Rev. B}\ }\textbf {\bibinfo {volume}
  {86}},\ \bibinfo {pages} {085119} (\bibinfo {year} {2012})}\BibitemShut
  {NoStop}%
\bibitem [{SM()}]{SM}%
  \BibitemOpen
  \href@noop {} {}\bibinfo {note} {See supplemental material for concrete
  examples and explanations of subspace separation and general construction
  method for block diagonalization, which includes Ref [51].}\BibitemShut
  {Stop}%
\bibitem [{\citenamefont {Meyer}(2000)}]{meyer_matrix_2000}%
  \BibitemOpen
  \bibfield  {author} {\bibinfo {author} {\bibfnamefont {C.~D.}\ \bibnamefont
  {Meyer}},\ }\href@noop {} {\emph {\bibinfo {title} {Matrix {Analysis} and
  {Applied} {Linear} {Algebra}}}}\ (\bibinfo  {publisher} {SIAM},\ \bibinfo
  {year} {2000})\BibitemShut {NoStop}%
\bibitem [{\citenamefont {Martin}\ \emph {et~al.}(2011)\citenamefont {Martin},
  \citenamefont {Di~Giuseppe}, \citenamefont {Perez-Leija}, \citenamefont
  {Keil}, \citenamefont {Dreisow}, \citenamefont {Heinrich}, \citenamefont
  {Nolte}, \citenamefont {Szameit}, \citenamefont {Abouraddy}, \citenamefont
  {Christodoulides},\ and\ \citenamefont {Saleh}}]{martin_anderson_2011}%
  \BibitemOpen
  \bibfield  {author} {\bibinfo {author} {\bibfnamefont {L.}~\bibnamefont
  {Martin}}, \bibinfo {author} {\bibfnamefont {G.}~\bibnamefont {Di~Giuseppe}},
  \bibinfo {author} {\bibfnamefont {A.}~\bibnamefont {Perez-Leija}}, \bibinfo
  {author} {\bibfnamefont {R.}~\bibnamefont {Keil}}, \bibinfo {author}
  {\bibfnamefont {F.}~\bibnamefont {Dreisow}}, \bibinfo {author} {\bibfnamefont
  {M.}~\bibnamefont {Heinrich}}, \bibinfo {author} {\bibfnamefont
  {S.}~\bibnamefont {Nolte}}, \bibinfo {author} {\bibfnamefont
  {A.}~\bibnamefont {Szameit}}, \bibinfo {author} {\bibfnamefont {A.~F.}\
  \bibnamefont {Abouraddy}}, \bibinfo {author} {\bibfnamefont {D.~N.}\
  \bibnamefont {Christodoulides}}, \ and\ \bibinfo {author} {\bibfnamefont
  {B.~E.~A.}\ \bibnamefont {Saleh}},\ }\href {\doibase 10.1364/OE.19.013636}
  {\bibfield  {journal} {\bibinfo  {journal} {Opt. Express}\ }\textbf {\bibinfo
  {volume} {19}},\ \bibinfo {pages} {13636} (\bibinfo {year}
  {2011})}\BibitemShut {NoStop}%
\end{thebibliography}

%

\end{document}